\documentstyle[preprint,aps,epsf]{revtex}

\newcommand {\bea}{\begin{eqnarray}}
\newcommand {\eea}{\end{eqnarray}}
\newcommand {\be}{\begin{equation}}
\newcommand {\ee}{\end{equation}}
\newcommand {\qslash}{q\!\!\!/}
\newcommand {\muslash}{\mu\!\!\!/}

\begin{document}

\preprint{SUNY-NTG-00-13}

\title{Quark Hadron Continuity in QCD with one Flavor}

\author{Thomas Sch\"afer}

\address{Department of Physics, SUNY Stony Brook,
Stony Brook, NY 11794\\ and\\
Riken-BNL Research Center, Brookhaven National 
Laboratory, Upton, NY 11973}

\maketitle

\begin{abstract}
  We study QCD with one flavor at finite baryon density. 
In the limit of very high baryon density the system is 
expected to be a color superconductor. In the case of one
flavor, the order parameter is in a $\bar 3$ of color and
has total angular momentum one. We show that in weak 
coupling perturbation theory, the energetically preferred
phase exhibits ``color-spin-locking'', i.e. the color 
and spin direction of the condensate are aligned. We 
discuss the properties of this phase and argue that it
shares important features of the hadronic phase at low 
density. In particular, we find an unbroken rotational
symmetry, spin 3/2 quasiparticles, and an unusual 
mechanism for quark-anti-quark condensation. Our results 
are relevant to three flavor QCD in the regime where the 
strange quark mass is bigger than the critical value for 
color-flavor-locking. We find that the gaps in this
case are on the order of one MeV.

\end{abstract}

\newpage

\section{Introduction}
\label{sec_int}

  The behavior of hadronic matter in the regime of very high 
baryon density but small temperature has attracted a lot 
of interest recently. It was realized a long time ago that 
asymptotic freedom combined with the presence of a Fermi surface 
implies that high density quark matter is a superconductor
\cite{Frau_78,Barrois:1977xd,Bar_79,Bailin:1984bm}. More 
recently, it was pointed out that the corresponding gaps
can be quite large, on the order of $\Delta\simeq 100$ MeV 
\cite{Alford:1998zt,Rapp:1998zu}. It was also realized
that the phase structure is quite rich, and that matter 
at very high density exhibits a wealth of non-perturbative 
phenomena, such as a mass gap and chiral symmetry breaking, 
in a regime where the coupling is weak and systematic 
calculations are possible \cite{Alford:1999mk,Schafer:1999ef}.

  The structure of the superconducting state depends 
sensitively on the number of quark flavors and their 
masses \cite{Alford:1999pa,Schafer:1999pb,Schafer:1999fe,Pisarski:1999gq}.
For two light flavors the dominant order parameter pairs
up and down quarks in a color anti-symmetric wave function. 
The condensate is a flavor singlet and breaks the gauge
symmetry $SU(3)\to SU(2)$. The up and down quarks that are 
singlets under the residual gauge symmetry remain gapless. 
In the case of three light flavors the pair condensate involves 
the coupling of color and flavor degrees of freedom, 
color-flavor-locking \cite{Alford:1999mk,Srednicki:1981cu}. 
Both the color and flavor symmetries are broken, but a
vector-like combination of the two remains a symmetry. 
This implies, among other things, that all gluons acquire 
a mass and that chiral symmetry is broken. In addition to
that, the spectrum of low-lying states bears an uncanny 
resemblance to what is expected, on phenomenological
grounds, for three flavor QCD at low baryon density. 
This has led us to conjecture that the hyperon matter
phase at low density might be continuously connected to 
quark matter at high density, without any phase transition
\cite{Schafer:1999ef}.

  If the strange quark mass is included, the structure 
of the ground state depends on the relative magnitude of 
the gap, the strange quark mass, and the Fermi momentum
\cite{Alford:1999pa,Schafer:1999pb}. If $\Delta>m_s^2/
(2p_F)$ all flavors participate and the system exhibits 
color-flavor-locking. If the gap is smaller then pairing 
only takes place in the up-down sector. In this
case, we expect the strange quarks to form an independent
superfluid \cite{Bailin:1984bm,Schafer:1999pb}. It is this
state we wish to study in more detail in this work. 

  At moderate densities, instantons play an important role
in determining the pairing gap for light quarks
\cite{Alford:1998zt,Rapp:1998zu,Carter:1999ji,Rapp:2000qa}. 
But instantons do not contribute to the scattering amplitude 
for two strange quarks. In the following, we will therefore 
focus on perturbative interactions \cite{Bar_79,Bailin:1984bm,Son:1999uk,Schafer:1999jg,Pisarski:2000tv,Hong:2000fh,Brown:1999aq}. This has 
the added advantage that our results 
are rigorous in the limit of very large chemical potential. 
If two strange quarks are in a color anti-symmetric wave 
function their combined spin and spatial wave function cannot
be antisymmetric. This means that pairing between strange quarks
has to involve total angular momentum one or greater. We 
will see that the perturbative one gluon exchange interaction 
is attractive for color anti-symmetric Cooper pairs in both 
the spin 0 and spin 1 states.

  Quark superfluids with total angular momentum one have
been studied using renormalization group methods 
\cite{Evans:1999ek,Evans:1999nf,Schafer:1999na,Hsu:1999mp},
perturbative QCD 
\cite{Bailin:1984bm,Pisarski:2000tv,Brown:1999yd,Iwasaki:1995ij}
and Nambu-Jona-Lasinio models \cite{Hosek:1998un}. In the
present work we wish to present a detailed investigation
of the phase structure and of the symmetries. For this 
purpose, we will not only consider the realistic case of 
three flavor matter with $m_s>\sqrt{2p_F\Delta}$, but also 
the academic situation of one flavor QCD at large chemical 
potential. In particular, we shall argue that one flavor 
QCD provides a new and interesting realization of the 
concept of quark-hadron continuity. 

 The paper is organized as follows. In section \ref{sec_nonrel}
we study the phase structure of one flavor QCD in the 
non-relativistic limit. We show that the stable phase
exhibits color-spin-locking, and discuss the symmetries
of this phase in section \ref{sec_csl}. In sections 
\ref{sec_rel} and \ref{sec_csl_rel} we study the 
phase structure in the ultra-relativistic limit. 
We conclude in section \ref{sec_con}.

\section{Phase Structure in the Non-Relativistic Limit}
\label{sec_nonrel}

  In practice we are mostly interested in the phase 
structure of superfluid strange quark matter. In this 
case, we expect that $p_F>m_s$ and the strange quark mass 
can be treated as a perturbation. Nevertheless, it is also 
interesting to consider the opposite limit $m>p_F$. The 
non-relativistic limit simplifies the calculation and we 
shall study it first. In the non-relativistic limit, 
the QCD lagrangian simplifies as
\be 
{\cal L} = \psi^\dagger\left( p_0+\mu -\frac{p^2}{2M}
 +gA_0 \right)\psi - \frac{1}{4}G^a_{\mu\nu}G^{a\,\mu\nu}.
\ee
We are interested in the behavior of one-flavor matter 
with $M>p_F>\Lambda_{QCD}$. This means that the density 
is sufficiently large to justify the perturbative 
treatment, but not so large as to invalidate the 
non-relativistic limit. The dominant interaction 
between quarks is given by the Coulomb force. This
interaction is attractive between quarks in a color
anti-symmetric $\bar 3$ state. In a
Fermi liquid, this attractive interaction will 
lead to an instability. We shall assume that this
instability is resolved by the formation of a 
condensate of quark pairs in a color $\bar 3$
state. 

 The Pauli principle requires that the wave function
of two identical quarks is antisymmetric. In QCD
with only one flavor this implies that color $\bar 3$
pairs cannot condense in a total angular momentum 
zero state. The obvious alternative is to consider
order parameters with total angular momentum one. 
In the non-relativistic limit there are only two
possibilities
\be
\label{qq_nr}
 (\phi_{s=1})_i^a = \psi i\sigma_2\sigma_i\lambda_A^a\psi,
 \hspace{1.5cm}
 (\phi_{l=1})_i^a = \psi i\sigma_2\hat{q}_i\lambda_A^a\psi.
\ee
Here, we have introduced a vector notation for the 
anti-symmetric Gell-Mann matrices $\vec{\lambda}_A=
(\lambda_2,\lambda_5,\lambda_7)$. The two order parameters 
are independent, because in the non-relativistic limit spin 
and orbital angular momentum are separately conserved.

 In order to derive a gap equation for the order parameters
defined in (\ref{qq_nr}) we follow the usual Nambu-Gorkov
procedure and introduce a bispinor $\Psi=(\psi,\psi_c)$
with $\psi_c =i\sigma_2\psi^\dagger$. In this basis, the 
fermionic part of the action becomes
\be
\label{S-1_nr}
 S^{-1}(q) = \left(\begin{array}{cc}
 q_0-\omega_q &  \Delta  \\
 \Delta  & q_0+\omega_q
\end{array}\right),
\ee
where we have defined $\omega_q=q^2/(2M)-\mu$. The interaction
vertex is a diagonal matrix $\Gamma^a={\rm diag}(\lambda^a/2,
(\lambda^a)^T/2)$. The Nambu-Gorkov matrix (\ref{S-1_nr}) is 
easily inverted to give the normal and anomalous components
of the quark propagator. The anomalous propagator is 
\be
 S_{21}(q) = \frac{\Delta}{q_0^2-\omega_q^2-\Delta^2},
\ee
where we have to keep in mind that $\Delta$ is a color-spin
matrix. The gap equation now follows from the Dyson-Schwinger
equation
\be
\label{ds_nr}
 \Sigma(k) = -ig^2\int\frac{d^4q}{(2\pi)^4} \Gamma^a S(q) \Gamma^b
 D^{ab}(q-k),
\ee
where $\Sigma(k)=-(S^{-1}(k)-S_0^{-1}(k))$ is the proper self energy
and $D^{ab}(q-k)$ is the Coulomb gluon propagator. 
 
  Before we proceed, we have to specify the color-spin structure
of the gap matrix. The gap is a $3\times 3$ matrix which transforms
as $\Delta\to U\Delta R$ under gauge transformations $U\in SU(3)$
and rotations $R\in SO(3)$. This is similar to the situation in 
liquid $^3$He \cite{VW_90} and in the color superconducting phase
of $N_f=3$ QCD \cite{Alford:1999mk}. In liquid $^3$He the order 
parameter describes the coupling of the nuclear spin to the 
orbital angular momentum of the pair. In the case of three flavor
QCD the order parameter specifies the coupling between the color
and flavor wave functions of the pair. While the order parameters
in the three cases look similar, the symmetries involved are not 
the same. Liquid $^3$He is characterized by a global $SO(3)\times 
SO(3)\times U(1)$ symmetry, high density QCD with one flavor by 
$SU(3)_c\times SO(3)\times U(1)$, and $N_f=3$ QCD by $SU(3)_c
\times SU(3)_L\times SU(3)_R\times U(1)$. The fact that the 
symmetry groups are not the same implies, among other things,
that the number of independent components of the order parameter 
is not the same. The same statement applies to the number of 
independent structures in the Landau-Ginzburg functional. We
find that the number of independent quartic terms in the 
Landau-Ginzburg functional is five in the case of $^3$He, 
three in the case of one flavor QCD, and two in the case
of three flavor QCD. 

  In any case, the preferred order parameter is governed by
dynamics, not symmetry, and has to determined for each system
separately. In the following, we shall consider the following
order parameters\footnote{In the context of color-flavor-locking,
it is sometimes argued that one can make use of color and flavor 
symmetries in order to restrict the possible order parameters to 
diagonal matrices. This is not necessarily correct, because there 
is no reason to exclude non-hermitean order parameters. An example 
is the A-phase of liquid $^3$He, which is one of the stable
phases of $^3$He at zero magnetic field.}
\be
\begin{array}{lcll}
 \Delta^a_i &=& \Delta \delta^a_i \hspace{1.5cm} & 
 {\rm B-phase\; (CSL)}, \\
 \Delta^a_i &=& \Delta \delta^{a3}\delta_{i3}  &
 {\rm polar\; phase\; (2SC)}, \\
 \Delta^a_i &=& \Delta \delta^{a3}(\delta_{i1}+i\delta_{i2}) & 
 {\rm A-phase }, \\
 \Delta^a_i &=& \Delta (\delta^{a1}\delta_{i1}+\delta^{a2}\delta_{i2})  &
 {\rm planar\; phase}, \\
\end{array}
\ee
which correspond to the ``inert'' phases of liquid $^3$He. These
phases are characterized by having the largest residual symmetry
groups, and by their stability under small perturbations that are 
consistent with the residual symmetry. There are three additional
inert phases in liquid $^3$He, the $A_1$, $B_1$ and $\beta$ 
phases \cite{VW_90}, but they do not lead to a gap on the Fermi
surface in the case of $N_f=1$ QCD. 

 In the following, we shall outline the calculation in the case
of the B-phase of the $s=1$ order parameter (\ref{qq_nr}). We 
shall refer to this phase as the color-spin-locked state, in 
analogy with the color-flavor-locked phase of $N_f=3$ QCD. The 
determination of the gap in the other phases proceeds along 
similar lines and we will briefly summarize the results below. 
In order to determine the anomalous propagator we have to 
diagonalize the color-spin structure of the gap matrix. In 
the case of color-spin locking this can be achieved using 
the observation that $(\vec{\sigma}\cdot\vec{\lambda}_A)^2=
2-(\vec{\sigma}\cdot\vec{\lambda}_A)$. This implies that the 
eigenvalues of $(\vec{\sigma}\cdot\vec{\lambda}_A)$ are 1 and 
$-2$. It is useful to introduce the corresponding projection 
operators
\bea
\label{P_1/2}
 P_{1/2} &=& \frac{1}{3}\Big(1-\vec{\sigma}\cdot
                      \vec{\lambda}_A \Big) \\
\label{P_3/2}
 P_{3/2} &=& \frac{2}{3}\Big(1+\frac{1}{2}\vec{\sigma}\cdot
                      \vec{\lambda}_A \Big) .
\eea
As we will explain in the next section, the subscript $g$
denotes the grand spin of the eigenstate. In particular, we
have ${\rm tr}(P_g)=(2g+1)$, so the degeneracies of the 
eigenvalues 1 and $-2$ are 4 and 2, respectively. It is 
now straightforward to determine the anomalous quark 
propagator. We find
\be 
\label{s21_nr_csl}
 S_{21}(q) = \frac{\Delta P_{3/2}}{q_0^2-\omega_q^2-\Delta^2}
   + \frac{-2\Delta P_{1/2}}{q_0^2-\omega_q^2-(2\Delta)^2}.
\ee
Using the explicit form of the projection operators, we 
can calculate the color factors for the two terms in the 
propagator. We find
\bea
\left(\frac{\lambda^a}{2}\right)P_{1/2}
 \left(\frac{\lambda^a}{2}\right)^T &=& 
-\frac{1}{3}P_{1/2} +\frac{1}{3} P_{3/2} \\
\left(\frac{\lambda^a}{2}\right)P_{3/2}
 \left(\frac{\lambda^a}{2}\right)^T &=& 
 \;\;\frac{2}{3}P_{1/2} .
\eea
The gap matrix is proportional to $P_{3/2}-2P_{1/2}$. 
For the gap equation to close, the right hand side of 
(\ref{ds_nr}) also has to be proportional to the same
combination of projectors. In the weak coupling limit
$\Delta\ll\mu$ we can neglect the difference of the 
gaps in the denominator of (\ref{s21_nr_csl}) and 
the gap equation indeed closes. At stronger coupling, 
the $g=1/2$ and $g=3/2$ gaps are independent, and 
there is a small admixture of a spin singlet, color
symmetric gap. This is analogous to the situation
in the color-flavor locked phase. Putting everything 
together, we find the following gap equation
\be
\label{ds_nr_csl}
\Delta(k) = \frac{2g^2}{9} \int\frac{d^4q}{(2\pi)^4}
 \left\{ \frac{\Delta(q)}{q_0^2-\omega_q^2-\Delta^2(q)}
   + \frac{2\Delta(q)}{q_0^2-\omega_q^2-(2\Delta(q))^2}
 \right\} D(q-k).
\ee
We take $D(p)$ to be a screened Coulomb propagator 
$D(p)=1/(\vec{p}^2+m_D^2)$ where $m_D^2=g^2\mu p_F/
(2\pi^2)$ is the Debye screening mass. In the weak coupling
limit the gap equation is dominated by momenta in the 
vicinity of the Fermi surface. In this case, there is 
no dependence on $k$ and we can approximate the gap
by a constant, $\Delta(k)\simeq \Delta$. We get 
\be
\Delta = \frac{2}{3}\frac{g^2}{8\pi^2}
 \log\left(\frac{4p_F^2}{m_D^2}\right)
 \frac{1}{3} \int_0^{\Lambda} d\epsilon\, 
 \Bigg\{ \frac{\Delta}{\sqrt{\epsilon^2+\Delta^2}}
     +  \frac{2\Delta}{\sqrt{\epsilon^2+(2\Delta)^2}}
 \Bigg\},
\ee
where we have introduced a cutoff $\Lambda$. 
This equation is easily solved 
\be
\Delta_{csl} = 2^{-2/3}(2\Lambda)\exp(-1/G), \hspace{1cm}
 G =  \frac{2}{3}\frac{g^2}{8\pi^2}
 \log\left(\frac{4p_F^2}{m_D^2}\right).
\ee
The factor $2^{-2/3}$ is due to the fact that the $g=\frac{1}
{2}$ and $g=\frac{3}{2}$ gaps are not equal. There is a similar 
factor $2^{-1/3}$ in the case of the color-flavor-locked
phase in $N_f=3$ QCD \cite{Schafer:1999fe}. We can 
now repeat this calculation for the other phases.
The main ingredient is the spectrum of gap matrix.
We find
\be
\begin{array}{lll}
{\rm polar\;phase}  & 
\Delta^\dagger\Delta = (\sigma_3\lambda_A^3)^2 & 
\lambda^2 = \left\{1\;(d\!=\!4),0\;(d\!=\!2)\right\}, \\
{\rm planar\;phase} & 
\Delta^\dagger\Delta = (\sigma_1\lambda_A^1+\sigma_2\lambda_A^2)^2 & 
\lambda^2 = \left\{2\;(d\!=\!4),0\;(d\!=\!2)\right\}, \\
{\rm A-phase} & 
\Delta^\dagger\Delta = (\sigma^+\lambda_A^3)(\sigma^-\lambda_A^3) & 
\lambda^2 = \left\{1\;(d\!=\!2),0\;(d\!=\!4)\right\}
\end{array}
\ee
where $d$ indicates the degeneracy of the eigenvalue. We observe
that the color-spin-locked phase is the only phase in which all
excitations are gapped. The gaps in the other phases are easily
calculated. We find
\be
 \Delta({\rm polar})=\Delta({\rm planar})=
 \Delta({\rm A-phase}) = 2^{2/3}\Delta({\rm CSL}),
\ee
which is a simple consequence of the fact that in the A, polar,
and planar phases all non-zero gaps are equal. Even though the 
gap in the CSL phase is smaller than the gap in the other 
phases, the critical temperature is not. The gap in the CSL
phase is suppressed because the $g=\frac{1}{2},\frac{3}{2}$
gaps are not equal. But $T_c$ is determined by the solution
of the finite temperature gap equation in the limit $\Delta
\to 0$, and does not depend on the spectrum of the gap 
matrix. 

 So far, we have only discussed the solution of the gap equation
in the different phases. The stable phase is determined by the 
condition that the thermodynamic potential is minimal. The grand
potential can be calculated from 
\be 
\label{F}
\Omega = \frac{1}{2} \int\frac{d^4q}{(2\pi)^4}
 \left\{ {\rm tr}\log\left(S_0^{-1}(q)S(q)\right)
   -{\rm tr}\left(S(q)\Sigma(q)\right) \right\}.
\ee
The traces can be calculated using the representation of the 
propagator in terms of projection operators (\ref{s21_nr_csl}). 
In the color-spin-locked phase, we find
\be 
\Omega = -\frac{\mu p_F}{2\pi^2}12\Delta^2_{csl}
  \log\left(\frac{2\mu}{\Delta_{csl}}\right).
\ee
In the polar, planar and A-phase the gap is bigger, but the number 
of condensed states is smaller. We find
\be
\label{f_comp}
\Omega({\rm polar})=\Omega({\rm A-phase}) = 
 \frac{2^{4/3}}{3} \Omega({\rm CSL}),
\ee
and $\Omega({\rm planar})=\frac{1}{2} \Omega({\rm polar})$. This 
result shows that the color-spin-locked phase is indeed favored, 
but only by a very small amount, $3\cdot 2^{-4/3}\simeq 1.2$. 

 Finally, we study the phases of the $l=1$ condensate $\langle
\psi\sigma_2\hat{q}_i\lambda_A^a\psi\rangle$. The gap depends
on the matrix $(\vec{d}(\hat{q})\cdot\vec{\lambda}_A)$ where
$d^a=\Delta^a_{\,i}\hat{q}_i$. The eigenvalues of $(\vec{d}
\cdot\vec{\lambda}_A)$ can be found from the fact that 
$(\hat{d}\cdot\vec{\lambda}_A)^3=(\hat{d}\cdot\vec{\lambda}_A)$.
This means that $\lambda=\{|\vec{d}|,-|\vec{d}|,0\}$, so that
there are always two gapless excitations, independent of 
the structure of $\vec{d}$. In the B-phase $\vec{d}=\hat{q}$, 
and the gap is isotropic. There are four gapped and two gapless 
modes. Even though the gap function is isotropic, the gap 
equation contains extra factors of $\cos(\theta)$ and the 
gap is reduced compared to the result in the polar phase 
of the $s=1$ order parameter. We will study these suppression 
factors in more detail in section \ref{sec_rel}. In the other 
phases the gap is no longer isotropic. In the polar phase, for 
example, we have $\vec{d}=\hat{e}_3$ and the gap behaves as 
$\Delta(\hat{q})\sim\cos(\theta)$. Because the gap is not 
isotropic, both the gap and the condensation energy are 
suppressed with respect to the B-phase.

\section{Symmetries of the Color-Spin-Locked Phase}
\label{sec_csl}

  The results of the previous section show that, in weak 
coupling and in the limit that the quark mass is large, high 
density QCD with one quark flavor exhibits color-spin-locking. 
In this section we wish to discuss some of the properties of t
he color-spin-locked phase. We will also contrast these properties 
with our expectations for the behavior of one flavor QCD at low 
density. 

 In the A-phase the $SO(3)\times SU(3)\times U(1)$ symmetry
is broken to $U(1)\times SU(2)$. Here, $U(1)$ is the residual
rotational symmetry and $SU(2)$ is the unbroken part of the 
gauge group. In the color-spin-locked phase the original 
rotational and gauge symmetries are completely broken, 
but there is a new $SO(3)$ invariance which is generated by a 
combination of the original $SO(3)$ generators and the $SU(3)$ 
color generators\footnote{Our discussion here applies to the 
idealized case of one flavor QCD. In QCD with three flavors
but separate pairing in the up-down and strange quark sectors
the color $SU(3)$ is broken to $SU(2)$ by the primary
condensate. This means that the color-spin-locked phase of
the strange superfluid in $N_f=3$ QCD cannot have exact rotational
invariance.}. Consider the ``grand spin'' generators
\be 
\vec{G} = \frac{\vec{\sigma}}{2} +\vec{\lambda}_A.
\ee
We can verify that the operators satisfy $SO(3)$ 
commutation relations
\be
 [G_i,G_j] = i\epsilon_{ijk} G_k,
\ee
and commute with the gap matrix in the color-spin-locked
phase
\be
 [\vec{G},(\vec{\sigma}\cdot\vec{\lambda}_A)]=0 .
\ee
This means that the color-spin-locked state is invariant 
with respect to rotations generated by $\vec{G}$. We can 
combine grand spin $\vec{G}$ and orbital angular momentum
$\vec{L}=\vec{r}\times\vec{p}$ to obtain the conserved
total angular momentum generator $\vec{J}=\vec{L}+\vec{G}$.
Away from the non-relativistic limit only $\vec{J}$ is 
conserved, not $\vec{L}$ and $\vec{G}$ separately. We 
will discuss this issue further in section \ref{sec_csl_rel}.

 Excitations in the color-spin-locked phase are characterized by 
their grand spin quantum numbers $g,g_3$. The quantum numbers of 
quasiparticles can be determined using 
\be
 {\vec{G}}^2 = \frac{11}{4} + 
 (\vec{\sigma}\cdot\vec{\lambda}_A).
\ee
We can now verify that the projectors $P_g$ defined in 
(\ref{P_1/2},\ref{P_3/2}) satisfy $\vec{G}^2P_g = g(g+1)P_g$. 
In the color-spin-locked phase there is a $g=\frac{3}{2}$
quartet of quasiparticles with gap $\Delta$, and a $g=
\frac{1}{2}$ doublet with gap $2\Delta$. All gluons 
acquire a mass via the Higgs mechanism. In the 
color-spin-locked phase the octet of gauge bosons
splits into a $g=1$ triplet and a $g=2$ quintet. 
If we couple orbital angular momentum to grand spin 
we find one $j=0$ state, two $j=1$ and $j=2$ states,
as well as one $j=3$ state. 

 In the color-spin-locked phase the $U(1)$ of baryon number
is spontaneously broken. As a consequence, the system 
exhibits superfluidity and the spectrum contains a massless
phonon. The order parameter is charged, and the photon 
acquires a mass by the Higgs mechanism. This also implies
that QCD with three flavors exhibits the Meissner effect
if the strange quark mass is larger than the critical  
mass for color-flavor-locking. Both the color-flavor-locked
phase and the phase with pairing in the up-down sector only
do not exhibit the Meissner effect. 

 We would now like to compare these results with our expectations
for the behavior of QCD with one flavor at low baryon density.
QCD with one flavor has a $U(1)_A$ chiral symmetry which is
broken by the anomaly. This means that there is no spontaneous
symmetry breaking, and that there are no Goldstone bosons. $N_f=1$
QCD has a large mass gap in all channels. The lowest dimension
operator with baryon number $B=1$ is a spin 3/2 current
\be
\label{eta_mu}
\eta_\mu = \epsilon^{abc}(q^aC\gamma_\mu q^b) q^c. 
\ee
This suggests that the lightest baryon has spin 3/2. This agrees 
with the expectation from the quark model. In order to construct
a color singlet state in which three quarks of the same flavor 
occupy an $s$-state the total spin has to be $3/2$. Finally, this
expectation is also borne out by phenomenology. In QCD, the lightest
$I=3/2$ baryon has spin $3/2$. The splitting between $s=1/2$ and
$s=3/2$ baryons with $I=3/2$ is about 400 MeV. The situation in 
the meson sector is less clear. For light quarks, phenomenology
suggests that both the scalar ($a_0$) and the pseudoscalar ($\eta'$)
are heavy, and the lightest state is a vector ($\omega$). For 
heavy quarks, on the other hand, the lightest quark-anti-quark
boundstate is a pseudoscalar, and the first excited state a vector. 
 
  Let us now turn to the effects of a non-zero chemical potential. 
Since the theory has a mass gap, there has to be a critical 
chemical potential $\mu_c\simeq M_B/3$ below which the baryon 
density is zero. Depending on whether nuclear matter is self-bound,
this transition is continuous or not. We have no information on the
interaction between two spin 3/2 baryons. If the lightest meson is 
a scalar, it is natural to assume that the $s$-wave scattering 
length is attractive. Even if this is not the case at very 
small density, repulsive interactions may get screened as the 
density is increased. In this case, we expect one flavor nuclear
matter to be a superconductor. A natural order parameter for
$s$-wave superconductivity is 
\be 
\label{phi_delta}
\phi = \langle \eta_\mu C\gamma_5 \eta_\mu \rangle .
\ee
This order parameter breaks the $U(1)$ of baryon number, and will 
lead to the appearance of a massless phonon. It also breaks the 
$U(1)$ of electromagnetism, and gives a mass to the photon. 

 If the density is very large we expect baryons to dissolve,
and it becomes natural to describe the system in terms of quarks. 
As we have seen, one-gluon exchange causes an instability near
the Fermi surface and the quark liquid is a color superconductor.
In the color-spin-locked phase rotational symmetry is unbroken,
and the only global symmetries that get broken are the $U(1)$ 
of baryon number and of electromagnetism. The gauge symmetry is 
completely broken and all colored excitations have a mass. In 
this sense, the system remains confined. This means that in 
terms of symmetries, the high density phase cannot be distinguished
from the low density phase. What is even more surprising is that
the spectrum of fermions is very similar to the low energy phase.
We saw that the spectrum contains spin 3/2 and spin 1/2 multiplets, 
where the spin 3/2 quasiparticles are lower in energy. This is 
the expected behavior in one flavor QCD at low density. These
observations suggest that the high density phase might be 
continuously connected to the low density phase, similar to what 
we suggested in the case of three flavor QCD \cite{Schafer:1999ef}.

\section{The Polar Phase in the Ultra-Relativistic Limit}
\label{sec_rel}

  We would now like to consider the opposite limit of massless,
ultra-relativistic quarks. In this section we shall study the
polar phase \cite{Pisarski:2000tv,Brown:1999yd}. More complicated
phases will be considered in the following section. The main new
ingredient in the relativistic limit is that the interaction 
preserves the chirality of the quarks. It is therefore useful 
to employ a chiral representation. In terms of left and right 
handed spinors $\psi_{L,R}$ the action becomes
\be
{\cal L}=
\bar\psi(\qslash+\muslash +gA\!\!\!/\,)\psi= 
\psi_R^\dagger ( q\cdot\sigma+\mu +gA\cdot\sigma )\psi_R +
\psi_L^\dagger ( q\cdot\bar\sigma+\mu +gA\cdot\bar\sigma )\psi_L,
\ee
where we have introduced $\sigma_\mu=(1,\vec{\sigma})$ and
$\bar\sigma_\mu=(1,-\vec{\sigma})$. There are two types 
of order parameters with total angular momentum one, depending 
on whether the condensate couples quarks of the same or opposite 
chirality. We begin with order parameters that connect quarks of 
the same chirality. We have
\bea
\label{qq-vec}
\psi C\gamma_5\hat{q}\lambda_2\psi &=& 
 -\psi_R i\sigma_2\hat{q}\lambda_2\psi_R 
 -\psi_L i\sigma_2\hat{q}\lambda_2\psi_L, \\
\label{qq-alp}
\psi C\vec{\alpha}\lambda_2\psi &=& 
 -\psi_R i\sigma_2\vec{\sigma}\lambda_2\psi_R 
 +\psi_L i\sigma_2\hat{\sigma}\lambda_2\psi_L,
\eea
where $\vec{\alpha}=\gamma_0\vec{\gamma}$ and we have selected 
a particular direction in color space. There are two additional 
order parameters with the opposite parity, $C\gamma_5\to C$ and 
$C\vec{\alpha}\to C\vec{\Sigma}$. In this case, the relative
sign between the $RR$ and $LL$ terms is flipped. As usual, 
perturbative interactions do not distinguish between order 
parameters of different parity. In the weak coupling limit, 
only states with the same chirality and helicity contribute. 
In order to make this manifest, we introduce the helicity 
projectors $H_\pm=\frac{1}{2}(1\pm\vec{\sigma}\cdot\hat{q})$. 
Using the fact that 
\be
 \psi_R \sigma_2 H_+ \vec{\sigma} H_+ \psi_R = 
 \psi_R \sigma_2 \hat{q} H_+ \psi_R
\ee
we see that, in the weak coupling limit, the second order 
parameter (\ref{qq-alp}) is not independent of the first
(\ref{qq-vec}). 

 In order to derive the gap equation we again consider the 
Dyson-Schwinger equation for the fermion self energy in the 
Nambu-Gorkov representation. We concentrate on right handed 
quarks and introduce the bispinor $\Psi=(\psi_R,\psi_{c,R})$
with $\psi_{c,R}=-i\sigma_2\psi_R^\dagger$. The inverse 
fermion propagator takes the form 
\be
\label{S-1RR}
S^{-1}(q) = \left(\begin{array}{cc}
 q\cdot\sigma +\mu &  \vec{\Delta}\cdot\hat{q}H_+ \\
 \vec{\Delta}^*\cdot\hat{q}H_+  & q\cdot\bar\sigma-\mu
\end{array}\right). 
\ee
The normal and anomalous components of the Nambu-Gorkov propagator 
are determined by the inverse of the matrix (\ref{S-1RR}). 
The anomalous propagator is given by 
\be 
S_{21} = -\frac{1}{q_-\cdot\bar\sigma} \;\Delta\; 
 \frac{1}{q_+\cdot\sigma-\Delta^\dagger (q_-\cdot\bar\sigma)^{-1}\Delta}
\ee
with $q_\pm=(q_0\pm\mu,\vec{q})$ and $\Delta=\vec{\Delta}\cdot\hat{q}H_+$. 
Except for the angular dependence of the gap parameter this propagator 
has the same form as the propagator in the spin 0 case. The corresponding 
gap equation has been discussed many times in the literature 
\cite{Son:1999uk,Schafer:1999jg,Pisarski:2000tv,Hong:2000fh}.  
Here, we simply quote the result
\bea
\label{sd_polar}
\Delta(k_0) &=& \frac{g^2}{12\pi^2} \int dq_0 \int d\cos(\theta)
 \frac{\cos(\theta)\Delta(q_0)}{\sqrt{q_0^2 +\cos(\theta)^2\Delta(q_0)^2}}
 \left\{ 
 \frac{\frac{1}{2}(1+\cos(\theta))}{1-\cos(\theta)+F^2/(2\mu^2)} 
 \right. \\
 & & \hspace{4cm}\left.\mbox{}+
 \frac{\frac{1}{2}(3-\cos(\theta))}{1-\cos(\theta)+G^2/(2\mu^2)}
 \right\}. \nonumber 
\eea
$G$ and $F$ are the magnetic and electric components of the 
gluon self energy. For $q_0\ll\vec{q}\ll\mu$ we have $F^2=
m_D^2$ and $G^2=\frac{\pi}{4}m_D^2x$ with $x=|q_0-k_0|/|\vec{q}
-\vec{k}|\simeq |q_0-k_0|/[\sqrt{2}\mu(1-\cos(\theta))]^{1/2}$.

 The gap equation is dominated by collinear scattering with 
$\cos(\theta)\simeq 1$. To leading order, we can solve the 
gap equation by setting $\cos(\theta)=1$ in the numerator. 
We also drop the angular dependence in the denominator of
the anomalous quark propagator. In this limit, the gap is
the same as for the spin zero case 
\be
 \Delta_0 = 512\pi^4(2/N_f)^{5/2}\mu g^{-5} 
 \exp\left(-\frac{3\pi^2}{\sqrt{2}g}\right),
\ee
where $N_f$ is the number of flavors that are active in 
determining the screening mass. Here, we only show the 
contribution of electric and magnetic gluon exchanges
to the pre-exponent. Additional contributions from 
the fermion self-energy were found in \cite{Brown:1999yd}. 
We can calculate corrections to the leading order result 
$\Delta_{l=1}=\Delta_{l=0}$ by expanding the numerator 
around $\cos(\theta)=1$. The correction term has 
no collinear singularity and the gluon self energy terms 
$F$ and $G$ can be neglected. We find $\Delta_{l=1}
=\exp(3c_1)\Delta_{l=0}$ with
\be
\label{l1_cor}
 c_1 = \frac{1}{2} \int \frac{d\cos(\theta)}{1-\cos(\theta)}\,
 \left\{\cos(\theta)\left(\frac{3}{2}-\frac{1}{2}\cos(\theta)\right)
     +\cos(\theta)\left(\frac{1}{2}+\frac{1}{2}\cos(\theta)\right)
     -2 \right\} = -2 .
\ee
This implies that $\Delta_{l=1}=\exp(-6)\Delta_{l=0}\simeq
0.004\Delta_{l=0}$ \cite{Brown:1999yd}, which shows that the 
angular momentum $l=1$ gap is strongly suppressed with respect
to the s-wave gap. While the natural scale of the s-wave gap 
is $\Delta=100$ MeV, the p-wave gap is expected to be less
than 1 MeV.

 We now come to superfluid order parameters that couple quarks 
of opposite chirality. In weak coupling, the only option is 
\be
\psi C\vec\gamma \lambda_2\psi = 
 -\psi_Li\sigma_2\vec{\sigma}\lambda_2\psi_R 
 -\psi_Ri\sigma_2\vec{\sigma}\lambda_2\psi_L .
\ee
In this case, the parity of the order parameter is fixed.
The order parameter $\psi C\gamma_5\vec{\gamma}\psi$ has
the opposite symmetry under the exchange of the two fermion
fields and cannot be color anti-symmetric. 

 We can derive the gap equation for $\psi C\vec{\Delta}
\cdot\vec{\gamma}\lambda_2\psi$ following the same steps
as in the spin zero case. We introduce the bispinor $\Psi=
(\psi_R,\psi_{c,L})$ with $\psi_{c,L}=i\sigma_2\psi_L^\dagger$. 
The inverse fermion propagator takes the form 
\be
\label{sinv}
S^{-1}(q) = \left(\begin{array}{cc}
 q\cdot\sigma +\mu &  H_+\vec{\Delta}\cdot\vec{\sigma}H_- \\
 H_-\vec{\Delta}\cdot\vec{\sigma}H_+  & q\cdot\sigma-\mu
\end{array}\right). 
\ee
The anomalous quark propagator is determined by the inverse
of this matrix. The gap equation is 
\bea
\label{s1_rel_gap}
\Delta(k_0) (\hat{\Delta}_\perp^k)^2 
 &=& \frac{g^2}{12\pi^2} \int dq_0 \int d\cos(\theta)
 \frac{\Delta(q_0)}
    {\sqrt{q_0^2 + \Delta(q_0)^2(\hat{\Delta}_\perp^q)^2 }}
    \frac{1}{2}\Bigg\{  \left( 
   1+ (\hat{k}\cdot\hat{\Delta})(\hat{q}\cdot\hat{\Delta}) \right)
  \left( 1+\hat{k}\cdot\hat{q} \right) \nonumber  \\  
 & & \hspace{2.5cm}\mbox{}
  - \left( \hat{k}\cdot\hat{\Delta}+\hat{q}\cdot\hat{\Delta} \right)^2
  -i\left( \hat{k}\cdot\hat{\Delta}+\hat{q}\cdot\hat{\Delta} \right)
    \hat{k}\cdot (\hat{\Delta}\times\hat{q}) \Bigg\} \nonumber \\
  & & \hspace{2.5cm}\mbox{}  
\cdot\left\{ 
 \frac{1}{1-\cos(\theta)+F^2/(2\mu^2)} +
 \frac{1}{1-\cos(\theta)+G^2/(2\mu^2)}
 \right\}.  
\eea
with $(\hat{\Delta}^k_\perp)^2=1-(\hat{k}\cdot\hat{\Delta})^2$.
We observe that in the absence of screening and damping, $F=G=0$, 
electric and magnetic gluon exchanges contribute equally to the 
spin 1 gap. We also note that the gap has a node if the direction 
of the order parameter is aligned with the pair momentum. This is due 
to the fact that the condensate connects quarks of opposite 
helicity. In order to produce a gap for a quark moving with 
momentum $\vec{p}$, the condensate has to flip the helicity 
of the quark. But this cannot happen if the spin of the 
condensate is parallel to $\vec{p}$. We can see this 
explicitly from the fact that
\be
\label{hel_mis}
 \bar\Delta\Delta = H_- \vec{\Delta}\cdot\vec{\sigma}H_+
  \vec{\Delta}\cdot \vec{\sigma} H_- 
 = H_-\left( (\vec{\Delta}\cdot\vec{\sigma})^2 -
   (\vec{\Delta}\cdot\hat{q})^2 \right) H_- .
\ee
If $\Delta\sim\hat{q}$ this expression vanishes, so 
there is no gap if $\hat{q}$ is parallel to $\vec{\Delta}$.

To leading order, we can solve the gap equation by evaluating 
the angular factors for $\hat{k}\cdot\hat{\Delta}=0$ and 
$\hat{k}\cdot\hat{q}=1$. In this case, the complicated 
matrix element in (\ref{s1_rel_gap}) reduces to the 
expression for the spin zero case \cite{Pisarski:2000tv}. 
The leading correction to this result can be determined
as in (\ref{l1_cor}). We find
\be
\label{s1_cor}
 c_{s=1} = \frac{1}{2}\int \frac{d\cos(\theta)}{1-\cos(\theta)}\,
 \left\{\cos(\theta)+\frac{1}{2}\cos(\theta)^2-\frac{3}{2}\right\} 
 = -\frac{3}{2}. 
\ee
This implies $\Delta^{LR}=\exp(-9/2)\Delta_0\simeq
0.01\Delta_{0}$, which is bigger than the $LL,RR$ gap 
by a factor $\sim 4.5$. 

 Finally, we have to consider the possibility that pairing
takes place both between quarks of the same and of opposite
chirality. In QCD with one flavor the chiral $U(1)_A$ symmetry
is anomalous, so there is no symmetry that prevents the two
condensates from mixing. But even if the chemical potential 
is infinitely large, and the effects of the anomaly can be
neglected, simultaneous pairing may still take place. As we
shall see, by combining the two order parameters we can 
obtain a gap with enhanced symmetries, and simultaneous
pairing could be energetically favored. In the following
we will consider a gap matrix of the form
\be
\label{gap_c}
 \Delta = \Lambda_+\left(\vec{\Delta}_1\cdot\hat{p}+
  \vec{\Delta}_2\cdot\vec{\gamma}\right)\Lambda_+\lambda_2 ,
\ee
where $\Lambda_\pm=\frac{1}{2}(1\pm\vec{\alpha}\cdot\hat{p})$
projects on positive (negative) energy states. This is equivalent
to projecting on states with equal chirality and helicity, 
$\Lambda_+=P_RH_++P_LH_-$. The derivation of the gap equation
proceeds along the same lines as before, only that now we have
to include both chiralities. The propagator takes the form
\be
\label{s21_rel_c}
 S_{21} = -\frac{1}{\qslash-\muslash} \;\Delta\; 
 \frac{1}{(\qslash+\muslash)+\bar\Delta 
  (\qslash-\muslash)^{-1}\Delta}.
\ee
The quadratic form in the denominator determines the structure
of the gap. We find
\be
\label{polar_cc}
\bar{\Delta}\tilde{\Delta}\equiv
 \bar{\Delta}(\qslash-\muslash)^{-1}\Delta (\qslash-\muslash)
 = \left[(\vec{\Delta}_1\cdot\hat{q})^2
  + \left( \vec{\Delta}_2^2-(\vec{\Delta}_2\cdot\hat{q})^2\right)
 \right]\Lambda_-,
\ee
where we have assumed that there is no relative phase between
$\vec{\Delta}_1$ and $\vec{\Delta}_2$. If this is not the case,
equation (\ref{polar_cc}) contains interference terms $\sim
{\rm Im}(\Delta_1^*\Delta_2)$ with a more complicated chiral
structure. There is no general reason why such terms should
be absent. Indeed, we shall find that interference between the
$LL$ and $LR$ terms is important in the color-spin-locked phase.
Here, we neglect interference effects for simplicity. Equation
(\ref{polar_cc}) shows that the case $\vec{\Delta}_1=\vec{\Delta}_2
\equiv \vec{\Delta}$ is special, because the gap function is 
completely isotropic. Nevertheless, rotational invariance is 
still broken. The gap is straightforward to determine. Since
the two structures in (\ref{gap_c}) have different chirality, 
the one-gluon exchange matrix elements decouple. As a result,
the gap equation is just a linear combination of the gap
equations in the $LL$ and $LR$ case. To leading order, the
gap is again identical to the spin zero gap. Taking into 
account the angular dependence of the matrix elements, the
gap is suppressed by $\Delta^{LL+LR}=\exp(-5)\Delta_0$. This
result simply corresponds to an average of the correction
terms in the $LL$ and $LR$ channel. Because the gap is 
isotropic, the condensation energy is increased by a 
factor 3/2 over pure $LR$ pairing. This is not sufficient,
however, to overcome the bigger angular suppression factor.
We conclude that in the limit of massless quarks, pairing
in the polar phase is dominated by the $LR$ and $RL$ 
channels\footnote{To leading order in the coupling, 
condensates with an arbitrary linear combination of 
$LL$ and $LR$ diquarks are degenerate. The degeneracy 
is lifted once higher order corrections are included. 
At present, however, it is not clear whether the 
next-to-leading order calculation performed in the 
present work is complete.}.

\section{Color-Spin-Locking in the ultra-relativistic Limit}
\label{sec_csl_rel}

 In the polar phase the color and spin orientation of the 
condensate are completely uncorrelated. In this section, we
shall deal with the more complicated cases in which the two are 
entangled. In practice, we will only discuss the color-spin-locked
phase. The calculation in the A and planar phases is very similar,
and we have verified that, in weak coupling, they do no compete
with the color-spin-locked phase. As in the previous section, we 
have to deal with three different cases, pairing between quarks 
of the same chirality, pairing between different chiralities, or a 
combination of the two. 

 We begin with the simplest case, which is pairing between
quarks of the same chirality. The gap matrix has the form
$\Delta=(\hat{q}\cdot\vec{\lambda_A})\Lambda_+$. As 
in section \ref{sec_nonrel}, this matrix can be diagonalized
using the relation $\Delta^3=\Delta$. This gives the eigenvalues
$\pm 1$ and 0. Projectors on the non-zero eigenvalues are given
by $P_{\pm 1} = \frac{1}{2}(\pm 1 + \Delta)\Delta$. We observe
that, just as in the non-relativistic limit, the B-phase of 
the RR (or LL) order parameter is not fully gapped. The value
of the gap is easy to determine. Using the projectors introduced
above we observe that the gap equation reduces to the one we
found in the polar phase, equ. (\ref{sd_polar}). This implies
$\Delta^{LL}_{csl}=\Delta^{LL}_{pol}=\exp(-6)\Delta_0$. Even 
though the gap in the B-phase is equal to the one in the polar 
phase, the condensation energy is not. Because the gap in the 
B-phase is isotropic, the condensation energy is three times 
larger and the B-phase is energetically favored.

 The B-phase of the $(LR+RL)$ order parameter is more complicated.
As in section \ref{sec_rel} we concentrate on the $LR$ sector and 
consider the gap matrix 
\be
\label{lr_csl}
\Delta = H_- (\vec{\sigma}\cdot\vec{\lambda}_A) H_+ .
\ee 
The physical gaps are determined by the eigenvalues of 
\be 
 \bar\Delta\Delta = H_+ \left[ (\vec{\sigma}\cdot\vec{\lambda}_A)^2
    -(\hat{q}\cdot\vec{\lambda}_A)^2\right] H_+ .
\ee
From $(\bar\Delta\Delta)^2=2(\bar\Delta\Delta)$ it follows that 
$\lambda=2,0$. The corresponding projectors are $P_2=\frac{1}{2}
\bar\Delta\Delta$ and $P_0=1-\frac{1}{2}\bar\Delta\Delta$. This 
shows that the eigenvalue $\lambda=2$ has degeneracy 2, and the 
eigenvalue $\lambda=0$ has degeneracy 1. Of course, the actual 
degeneracies are 4 and 2, because of the trivial degeneracy $LR
\to RL$. In the non-relativistic limit $\psi C\vec{\gamma}\cdot
\vec{\lambda}_A\psi$ reduces to $\psi\sigma_2\vec{\sigma}\cdot
\vec{\lambda}_A\psi$. But while the non-relativistic order parameter 
$\psi\sigma_2\vec{\sigma}\cdot\vec{\lambda}_A\psi$ leads to a fully
gapped state, we find that in the ultra-relativistic limit two
modes remain gapless. This is related to the result that in the
polar phase the $LR+RL$ gap has a node on the Fermi surface,
see equation (\ref{hel_mis}). 

 This also means that, unlike the non-relativistic case
(\ref{ds_nr_csl}) the gap equation is not modified as compared
to the polar phase. We find $\Delta^{LR}_{csl}=\Delta^{LR}_{pol} 
= \exp(-9/2)\Delta_0$ where $\Delta_0$ is the spin zero gap. On 
the other hand, because the gap is isotropic, the condensation 
energy is increased compared to the polar phase by a factor $3/2$. 

 Finally, we have to consider the possibility that pairing takes 
place in both the $LL,RR$ and $LR,RL$ channels. In particular,
we would like to consider the gap matrix
\be
\label{gap_ccsl}
 \Delta = \Lambda_+ \left(\hat{q}+\vec{\gamma}\right)
 \cdot\vec{\lambda}_A\Lambda_+ .
\ee
We note that this order parameter has positive parity, and 
that the parity is fixed, even if only perturbative interactions
are taken into account. The physical gaps are determined by 
(c.f. equation \ref{polar_cc})
\be
\label{dd_ccsl}
\bar{\Delta}\tilde{\Delta} = 
 \Lambda_-\left[ (\vec{\alpha}\cdot\vec{\lambda}_A)^2
 +i\vec{\gamma}\cdot(\hat{q}\times\vec{\lambda}_A)\right]
 \Lambda_- .
\ee
We note that, similar to the polar case, the structure $(\hat{q}
\cdot\vec{\lambda}_A)$ has disappeared, but this time an interference 
term is present. We can now follow the standard procedure and determine 
the characteristic equation for $\bar\Delta\tilde\Delta$. We 
find $(\bar\Delta\tilde\Delta)^3-5(\bar\Delta\tilde\Delta)^2
+4(\bar\Delta\tilde\Delta)=0$ which leads to the eigenvalues
$\lambda=0,1,4$. The corresponding projectors are
\be
\label{ccsl_proj}
 P_0 = \Lambda_+, \hspace{0.5cm}
 P_{3/2} = \frac{4}{3} \Big(1-\frac{1}{4}(\bar\Delta\tilde\Delta)
       \Big)\Lambda_-, \hspace{0.5cm}
 P_{1/2} = -\frac{1}{3} \Big(1-(\bar\Delta\tilde\Delta)\Big)\Lambda_-,
\ee
where we follow the notation used in (\ref{P_1/2},\ref{P_3/2}). 
We note that in the weak coupling limit, all particles are gapped. 
The eigenvalue $\lambda=1$ has multiplicity 4, while $\lambda=4$
has multiplicity 2. This implies that the spectrum is identical 
to the one we found in the non-relativistic
color-spin-locked phase. As a consequence, the structure of the 
gap equation is also very similar to (\ref{ds_nr_csl}). We
find
\be
\Delta(k_0) = g^2\int\frac{d^4q}{(2\pi)^4} \left\{
 \frac{\Delta(q_0)M^{\mu\nu}_{3/2}}{q_0^2+\omega^2_q+(\Delta(q_0))^2}
 +\frac{\Delta(q_0)M^{\mu\nu}_{1/2}}{q_0^2+\omega^2_q+(2\Delta(q_0))^2}
 \right\} D_{\mu\nu}(q-k),
\ee
with $\omega^2_q=(q-\mu)^2$ and the matrix elements
\be
M^{\mu\nu}_g = \frac{1}{12}{\rm tr} 
 \Bigg(\gamma_\mu  \left(\frac{\lambda^a}{2}\right)
 \Lambda_-^q \vec{\lambda}_A\cdot\left(\hat{q}-\vec{\gamma}\right)
 \Lambda_-^q P_g \gamma_\nu \left(\frac{\lambda^a}{2}\right)^T
 \Lambda_+^k \vec{\lambda}_A\cdot\left(\hat{k}-\vec{\gamma}\right)
 \Lambda_+^k \Bigg)
\ee
where $\Lambda_\pm^q=\frac{1}{2}(1\pm\vec{\alpha}\cdot\hat{q})$,
$\Lambda_\pm^k=\frac{1}{2}(1\pm\vec{\alpha}\cdot\hat{k})$ and $P_g$
are the projectors defined in (\ref{ccsl_proj}). The structure of 
these matrix elements is quite complicated, but the traces simplify 
in the weak coupling limit. In this case, we can evaluate the matrix 
elements in the forward direction $\vec{q}\simeq\vec{k}$ and find
\bea
\label{csl_rel_gap}
\Delta(k_0)  &=& \frac{g^2}{12\pi^2} \int dq_0 \int d\cos(\theta)
 \left\{ \frac{1}{3}\frac{\Delta(q_0)}{\sqrt{q_0^2 + (\Delta(q_0))^2}}
    + \frac{2}{3}\frac{\Delta(q_0)}{\sqrt{q_0^2+(2\Delta(q_0))^2}}
 \right\} \\
  & & \hspace{3.5cm}\mbox{}  
\cdot\left\{ 
 \frac{1}{1-\cos(\theta)+F^2/(2\mu^2)} +
 \frac{1}{1-\cos(\theta)+G^2/(2\mu^2)}
 \right\}. \nonumber 
\eea
In leading order we can neglect the difference between the gaps 
and find $\Delta^{LL+LR}_{csl}=\Delta_0$. Taking the difference into 
account we get $\Delta^{LL+LR}_{csl}=2^{-2/3}\Delta_0$, as in the 
non-relativistic case. Again, the condensation energy is bigger as 
compared to the polar state, even though the gap is smaller. We can 
also determine the corrections which come from non-forward scattering. 
The calculation is identical to the one in the polar phase and we
find $\Delta^{LL+LR}_{csl}=2^{-2/3}\exp(-5)\Delta_0$. 

 The color-spin-locked state (\ref{dd_ccsl}) has a number of 
interesting properties. First we note that, as in the non-relativistic
limit, rotational invariance is unbroken, and the low-lying fermions
are organized into a spin 3/2 and a spin 1/2 multiplet. What is new
in the relativistic case is the fact that both the $U(1)_V$ and 
$U(1)_A$ symmetries are broken. The $U(1)_A$ symmetry is also broken 
in the polar phase, but there is an important difference here. Without 
the chirally odd interference term in (\ref{dd_ccsl}) the diquark
condensate does not induce a quark-anti-quark condensate $\langle
\bar\psi_L\psi_R\rangle$. In the color-spin-locked state, there is a
non-vanishing condensate
\be
\Sigma_{csl} = \langle \bar\psi \vec{\alpha}\cdot(\hat{q}\times
 \vec{\lambda}_A)\psi\rangle .
\ee
In the weak coupling limit we find $\Sigma_{csl}=-\frac{1}{6\pi^2}
\mu^3\log(2)$. The condensate is a scalar under rotations generated 
by the grand angular momentum operator. This means that is has the 
same symmetries as the quark condensate $\langle\bar\psi\psi\rangle$. 
Once higher order perturbative corrections are included, we expect 
the primary condensate $\Sigma_{csl}$ to induce a non-zero $\langle
\bar\psi\psi\rangle$ as well. As a result, there will be a non-zero 
quark condensate even at very large density, where instantons are 
exponentially suppressed. 

 We saw that the gap in the state (\ref{gap_ccsl}), which has 
an equal mixture of $LL$ and $LR$ components, is given by 
$\Delta^{LL+LR}_{csl}=2^{-2/3}\exp(-5)\Delta_0$. This is suppressed 
with respect to the gap in the color-spin-locked phase of the pure 
$LR$ order parameter, $\Delta^{LR}_{csl}=\exp(-4.5)\Delta_0$, and 
the larger number of condensed species is insufficient to overcome
this suppression. On the other hand, we saw that the spectrum in 
the $LL+LR$ state corresponds exactly to the spectrum in the 
non-relativistic limit. This suggests that, if the baryon density
is increased in $N_f=1$ QCD with massive quarks, the order parameter
evolves from the fully gapped $LL+LR$ state to the partially gapped
$LR$ state. In order to see whether this can happen continuously we
would like to study the spectrum for a general linear combination
of the $LL$ and $LR$ order parameters
\be 
\label{gap_cccsl}
 \Delta = \Lambda_+ \left(\cos(\beta)\hat{q}+\sin(\beta)\vec{\gamma}
 \right)\cdot\vec{\lambda}_A\Lambda_+ .
\ee
For $\beta=\pi/2$ this corresponds to the pure $LR$ order parameter
(\ref{lr_csl}), and for $\beta=\pi/4$ we get an equal mixture of
$LL$ and $LR$ as in (\ref{gap_ccsl}). For an arbitrary value of 
$\beta$ the spectrum of $\bar\Delta\tilde\Delta$ is given by
\be
\label{spec_cccsl}
 \lambda_1 = \cos(\beta)^2, \hspace{1cm}
 \lambda_{2,3} = \frac{1}{4} \left( 5 \pm \sqrt{2}\cos(\beta)
   \sqrt{9-7\cos(2\beta)} - 3\cos(2\beta)\right),
\ee
where all eigenvalues are doubly degenerate. We show the spectrum 
as a function of $\beta$ in Fig.1. We note that the spectrum 
is fully gapped for all values of $\beta$ except for $\beta=0,\pi/2$. 
There are three values of $\beta$, $\beta=0,\pi/4,\pi/2$, for which
two pairs of eigenvalues meet, and the degeneracy of the spectrum 
is enhanced. These correspond to the cases we already discussed in
detail, pure $LL$ and $LR$ pairing, and equal $LL$ and $LR$ pairing. 
To leading order in the coupling, the gap and the condensation 
energy are independent of the mixing angle $\beta$. Taking subleading
corrections into account, we found that in the non-relativistic limit 
the state corresponding to $\beta=\pi/4$ is favored. In the 
ultra-relativistic limit, the energetically preferred state
has $\beta=\pi/2$. We therefore conjecture that as a function
of $p_F/m$ the order parameter evolves from $\beta=\pi/4$ to
$\beta=\pi/2$.


\section{Conclusions}
\label{sec_con}

 In summary, we have studied QCD with one flavor at high baryon
density. Our results are relevant to QCD with three flavors in the 
case when the strange quark mass is bigger than the critical value 
for color-flavor-locking \cite{Alford:1999pa,Schafer:1999pb}. They 
also apply to the situation in two flavor QCD when the difference 
between the chemical potentials for up and down quarks is bigger 
than the gap \cite{Bedaque:1999nu,Son:2000xc}. In both cases there 
is no pairing between quarks of different flavors, and the 
possible phases are identical to those in one flavor QCD. We 
should note, however, that even if the pair condensate involves 
only a single flavor, there will still be some dependence on
the number flavors. This dependence arises from the $N_f$ 
dependence of the screening mass, and from higher order 
corrections that may couple condensates of different flavors. 

 In QCD with one flavor, as in QCD with two or more flavors, the 
Fermi surface is unstable with respect to the formation of 
color anti-symmetric Cooper pairs. However, because the wave
function of the pair cannot be anti-symmetric in flavor, the 
Cooper pairs have to have angular momentum one or greater. 
As a result, the magnitude of the gap is suppressed with 
respect to the spin zero gap in two flavor QCD. Using weak
coupling perturbation theory, we find $\Delta<\exp(-4.5)
\Delta_0$ where $\Delta_0$ is the spin zero gap. If the 
typical magnitude of the gap in two flavor QCD is 100 MeV,
we find that the one flavor gap is $\Delta\sim$ 1 MeV. 

 In one flavor QCD the order parameter is a spin-color matrix, 
and interesting phases can arise because of the possibility 
that color and spin degrees of freedom become entangled. The
situation is superficially similar to the phase structure of 
liquid $^3$He \cite{VW_90} and high density QCD with three flavors 
\cite{Schafer:1999fe,Shovkovy:1999mr,Evans:1999at}, but the
dynamics and the symmetries involved are different. Nevertheless,
as in BCS (or Eliashberg) studies of $N_f=3$ QCD or liquid $^3$He, 
we find that, in weak coupling perturbation theory, the B-phase is 
energetically favored. In analogy with the color-flavor-locked
phase of $N_f=3$ QCD we refer to this phase as color-spin-locked. 

 The situation is particularly simple in the non-relativistic 
limit. In this case, there is a unique groundstate, and the 
spectrum in the color-spin-locked phase is fully gapped. 
In the ultra-relativistic limit the situation is more complicated.
The order parameter exhibits color-spin-locking, but to leading
order in the coupling constant there is a continuous family of
states which differ by the mixing angle between the $\psi_R\psi_R-
\psi_L\psi_L$ and $\psi_R\psi_L+\psi_L\psi_R$ components of the 
order parameter. Except at special points, the spectrum is again
fully gapped. 

 In the color-spin-locked phase the original rotational symmetry
is broken, but there is an unbroken $SO(3)$ symmetry which is 
generated by a combination of the original angular momentum 
and color generators. The only non-anomalous symmetry which 
is broken in the color-spin-locked phase is the $U(1)$ of 
baryon number. This means that the global symmetries of the 
color-spin-locked phase agree with what we expect, on phenomenological
grounds, for one flavor QCD at low density. We also found that the
color-spin-locked phase has certain other features that are 
characteristic of $N_f=1$ QCD. In particular, we saw that the
color-spin-locked phase supports low energy spin 3/2 quasiparticles,
and that there is a mechanism for generating quark-anti-quark 
condensates. These observations lead us to conjecture that 
in one flavor QCD the low and high density phases are 
continuously connected. In the case of one flavor QCD this
suggestion is less radical than in the case of three flavors.
In particular, it is known that for sufficiently small values
of the quark mass there is no phase transition along the finite 
temperature axis \cite{Alexandrou:1999wv}. In this case, we
expect the only phase transition in the $T-\mu$ plane to 
be the nuclear onset transition.

 Acknowledgements: This work was supported in part by US DOE grant 
DE-FG-88ER40388. I would like to acknowledge the hospitality of the 
National Institute for Nuclear Theory in Seattle, where this work 
was completed.

\newpage\noindent

\begin{figure}
\begin{center}
\leavevmode
\vspace{1cm}
\epsfxsize=10cm
\epsffile{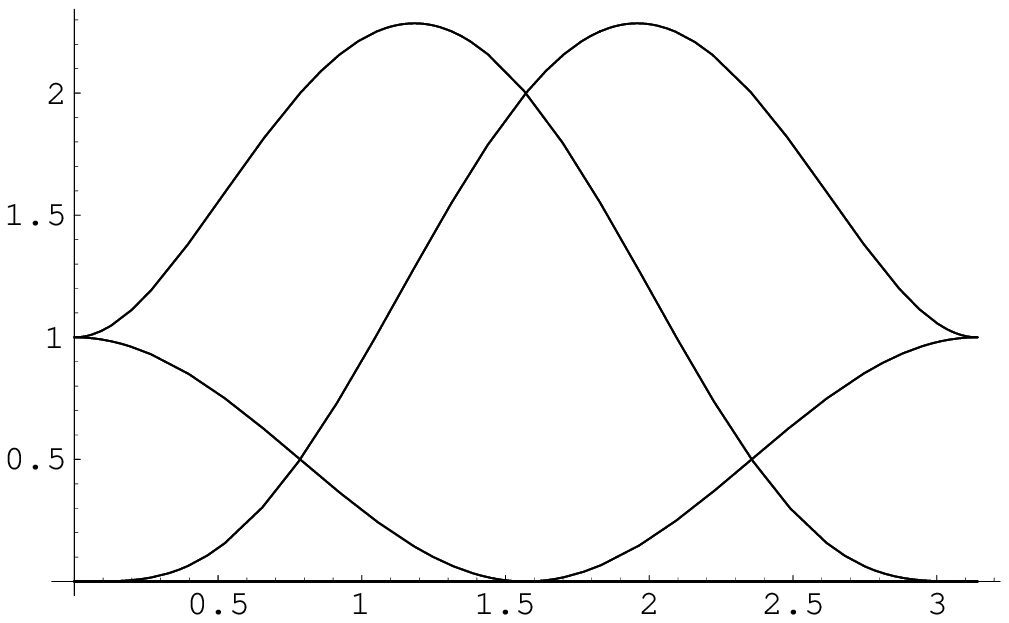}
\end{center}
\caption{Quasiparticle spectrum in the color-spin-locked phase
as a function of the mixing angle $\beta$ between the $LL$ and
$LR$ condensates, see equation (\ref{spec_cccsl}).}
\end{figure}


\newpage



\begin{thebibliography}{20}

\bibitem{Frau_78}
S. C. Frautschi,
Asymptotic freedom and color superconductivity in dense quark matter,
in: Proceedings of the
Workshop on Hadronic Matter at Extreme Energy Density, N. Cabibbo, Editor,
Erice, Italy (1978).

\bibitem{Barrois:1977xd}
B.~C.~Barrois,
Nucl.\ Phys.\  {\bf B129}, 390 (1977).
 
\bibitem{Bar_79}
F. Barrois, Nonperturbative effects in dense quark matter,
Ph.D. thesis, Caltech, UMI 79-04847-mc (microfiche).

\bibitem{Bailin:1984bm}
D.~Bailin and A.~Love,
Phys.\ Rept.\  {\bf 107}, 325 (1984).
 
\bibitem{Alford:1998zt}
M.~Alford, K.~Rajagopal and F.~Wilczek,
Phys.\ Lett.\  {\bf B422}, 247 (1998)
[hep-ph/9711395].
 
\bibitem{Rapp:1998zu}
R.~Rapp, T.~Sch{\"a}fer, E.~V.~Shuryak and M.~Velkovsky,
Phys.\ Rev.\ Lett.\  {\bf 81}, 53 (1998)
[hep-ph/9711396].
 
\bibitem{Alford:1999mk}
M.~Alford, K.~Rajagopal and F.~Wilczek,
Nucl.\ Phys.\  {\bf B537}, 443 (1999)
[hep-ph/9804403].
  
\bibitem{Schafer:1999ef}
T.~Sch{\"a}fer and F.~Wilczek,
Phys.\ Rev.\ Lett.\  {\bf 82}, 3956 (1999)
[hep-ph/9811473].

\bibitem{Alford:1999pa}
M.~Alford, J.~Berges and K.~Rajagopal,
Nucl.\ Phys.\  {\bf B558}, 219 (1999)
[hep-ph/9903502].
 
\bibitem{Schafer:1999pb}
T.~Sch{\"a}fer and F.~Wilczek,
Phys.\ Rev.\  {\bf D60}, 074014 (1999)
[hep-ph/9903503].
 
\bibitem{Schafer:1999fe}
T.~Sch{\"a}fer,
hep-ph/9909574.

\bibitem{Pisarski:1999gq}
R.~D.~Pisarski,
nucl-th/9912070.

\bibitem{Srednicki:1981cu}
M.~Srednicki and L.~Susskind,
Nucl.\ Phys.\  {\bf B187}, 93 (1981).

\bibitem{Carter:1999ji}
G.~W.~Carter and D.~Diakonov,
Phys.\ Rev.\  {\bf D60}, 016004 (1999)
[hep-ph/9812445].

\bibitem{Rapp:2000qa}
R.~Rapp, T.~Sch{\"a}fer, E.~V.~Shuryak and M.~Velkovsky,
Annals Phys.\  {\bf 280}, 35 (2000)
[hep-ph/9904353].
  
\bibitem{Son:1999uk}
D.~T.~Son,
Phys.\ Rev.\  {\bf D59}, 094019 (1999)
[hep-ph/9812287].
 
\bibitem{Schafer:1999jg}
T.~Sch{\"a}fer and F.~Wilczek,
Phys.\ Rev.\  {\bf D60}, 114033 (1999)
[hep-ph/9906512].
  
\bibitem{Pisarski:2000tv}
R.~D.~Pisarski and D.~H.~Rischke,
Phys.\ Rev.\  {\bf D61}, 074017 (2000)
[nucl-th/9910056].
  
\bibitem{Hong:2000fh}
D.~K.~Hong, V.~A.~Miransky, I.~A.~Shovkovy and L.~C.~Wijewardhana,
Phys.\ Rev.\  {\bf D61}, 056001 (2000)
[hep-ph/9906478].
 
\bibitem{Brown:1999aq}
W.~E.~Brown, J.~T.~Liu and H.~Ren,
hep-ph/9908248.
   
\bibitem{Evans:1999ek}
N.~Evans, S.~D.~Hsu and M.~Schwetz,
Nucl.\ Phys.\  {\bf B551}, 275 (1999)
[hep-ph/9808444].
 
\bibitem{Evans:1999nf}
N.~Evans, S.~D.~Hsu and M.~Schwetz,
Phys.\ Lett.\  {\bf B449}, 281 (1999)
[hep-ph/9810514].
 
\bibitem{Schafer:1999na}
T.~Sch{\"a}fer and F.~Wilczek,
Phys.\ Lett.\  {\bf B450}, 325 (1999)
[hep-ph/9810509].
 
\bibitem{Hsu:1999mp}
S.~D.~Hsu and M.~Schwetz,
hep-ph/9908310.
 
\bibitem{Brown:1999yd}
W.~E.~Brown, J.~T.~Liu and H.~Ren,
hep-ph/9912409.
    
\bibitem{Iwasaki:1995ij}
M.~Iwasaki and T.~Iwado,
Phys.\ Lett.\  {\bf B350}, 163 (1995).

\bibitem{Hosek:1998un}
J.~Hosek, 
hep-ph/9812516.
 
\bibitem{VW_90}
D. Vollhardt and P. W{\"o}lfe, 
{\it The superfluid phases of Helium 3}, 
Taylor and Francis (1990).
 
\bibitem{Bedaque:1999nu}
P.~Bedaque,
hep-ph/9910247.

\bibitem{Son:2000xc}
D.~T.~Son and M.~A.~Stephanov,
hep-ph/0005225.

\bibitem{Shovkovy:1999mr}
I.~A.~Shovkovy and L.~C.~Wijewardhana,
Phys.\ Lett.\  {\bf B470}, 189 (1999)
[hep-ph/9910225].
 
\bibitem{Evans:1999at}
N.~Evans, J.~Hormuzdiar, S.~D.~Hsu and M.~Schwetz,
hep-ph/9910313.

\bibitem{Alexandrou:1999wv}
C.~Alexandrou, A.~Borici, A.~Feo, P.~de Forcrand, A.~Galli, 
F.~Jegerlehner and T.~Takaishi, 
Phys.\ Rev.\ {\bf D60}, 034504 (1999)
[hep-lat/9811028].
 
\end{thebibliography}
\end{document}